# Report on the Twelfth United Nations/European Space Agency Workshop on Basic Space Science

## (Beijing, P.R. China, 24-28 May 2004)

Willem Wamsteker

European Space Agency, VILSPA Villafranca del Castillo Satellite Tracking Station, P.O. Box 50727, 28080 Madrid, Spain,

Hans J. Haubold

United Nations Office for Outer Space Affairs, Vienna International Centre, A-1400 Vienna, Austria, Hans.Haubold@unvienna.org

# I. Introduction

## A. Background and objectives

1. The Third United Nations Conference on the Exploration and Peaceful Uses of Outer Space (UNISPACE III) and the Vienna Declaration on Space and Human Development recommended that activities of the United Nations Programme on Space Applications promote collaborative participation among Member States at both the regional and international levels, emphasizing the development of knowledge and skills in developing countries.[i]

2. At its forty-sixth session, in 2003, the Committee on the Peaceful Uses of Outer Space endorsed the programme of workshops, training courses, symposiums and conferences planned for 2004.[ii] Subsequently, the General Assembly, in its resolution 58/89 of 9 December 2003, endorsed the United Nations Programme on Space Applications for 2004.

3. Pursuant to resolution 58/89 and in accordance with the recommendation of UNISPACE III, the Twelfth United Nations/European Space Agency Workshop on Basic Space Science was organized by the United Nations, the European Space Agency (ESA) and the Government of China in Beijing from 24 to 28 May 2004. The China National Space Administration (CNSA) hosted the Workshop on behalf of the Government of China.

4. The Workshop was the latest in a series of United Nations/ESA workshops on basic space science that had been organized for the benefit of developing countries in India (1991) and Sri Lanka (1996) for Asia and the Pacific (see A/AC.105/489 and A/AC.105/640); in Costa Rica (1992) and Honduras (1997) for Central America (see A/AC.105/530 and A/AC.105/682); in Colombia (1992) and Argentina (2002) for South America (see A/AC.105/530 and A/AC.105/784), in Nigeria (1993) and Mauritius (2001) for Africa (see A/AC.105/560/Add.1 and A/AC.105/766); in Egypt (1994) and Jordan (1999) for Western Asia (see A/AC.105/580 and A/AC.105/723); and in Germany (1996) and France (2000) for Europe (see A/AC.105/657 and A/AC.105/742). The workshops were co-organized by the Abdus Salam International Centre for Theoretical Physics, the Austrian Space Agency, the Centre national d'études spatiales of France, the Committee on Space Research (COSPAR), ESA, the German Space Agency (DLR), the Institute of Space and Astronautical Science of Japan, the International Astronomical Union (IAU), the National Aeronautics and Space Administration (NASA) of the United States of America, the National Astronomical Observatory of Japan, the Planetary Society and the United Nations.

5. The main objective of the Workshop was to provide a forum to highlight recent scientific and technical results obtained using major space-borne observatories for studies of the planets, stars and the far reaches of the universe. Such satellite missions constitute an impressive means of studying all aspects of basic space science from space as a complement to studies being done from the ground. The question of the large volumes of data generated by such missions was discussed in relation to changing research needs within the scientific community, as well as the question of how access to the important databases maintained by space agencies could be facilitated. The issues of data, research and education based on space missions were discussed, together with the relevance of such missions to those developing countries wishing to participate actively in the voyage of discovery through the solar system and the universe.

6. The present report was prepared for submission to the Committee on the Peaceful Uses of Outer Space at its forty-eighth session and to its Scientific and Technical Subcommittee at its forty-second session, in 2005.

### B. Programme

7. At the opening of the Workshop, introductory statements were made by representatives of the Ministry of Foreign Affairs of China, CNSA, the Chinese Academy of Sciences, ESA and the United Nations. The Workshop was divided into scientific sessions, each focusing on a specific issue. Presentations by invited speakers describing the status of their findings in research and education were followed by brief discussions. Fifty papers were presented by invited speakers from both developing and developed countries. Poster sessions and working group sessions provided an opportunity to focus on specific problems and projects in basic space science.

8. The session of the Workshop focused on: (a) astrophysical data systems, archives and distribution of knowledge; (b) virtual observatories; (c) non-extensive statistical mechanics and astrophysics; (d) ways and means to accelerate the development of basic space science; (e) concurrent design capabilities for the development of international space missions; (f) new opportunities for space missions in astrophysics and the solar system; (g) planetary exploration; and (h) preparations for the International Heliophysical Year.

### C. Attendance

9. Researchers and educators from developing and industrialized countries from all economic regions were invited by the United Nations, ESA and CNSA to participate in the Workshop. Participants held positions at universities, research institutions, observatories, national space agencies and international organizations and were involved in all the aspects of basic space science covered by the Workshop. Participants were selected on the basis of their scientific background and their experience with programmes and projects in which basic space science played a leading role.

10. Funds provided by the United Nations, ESA and CNSA were used to cover the travel, living and other costs of participants from developing countries. A total of 75 specialists in basic space science attended the Workshop.

11. The following 28 Member States were represented at the Workshop: Austria, Brazil, Canada, China, Ethiopia, France, Germany, India, Indonesia, Iraq, Italy, Japan, Malaysia, Mexico, Netherlands, Paraguay, Poland, Russian Federation, Singapore, South Africa, Spain, Syrian Arab Republic, United Arab Emirates, United Kingdom of Great Britain and Northern Ireland, Uruguay, United States, Viet Nam and Yemen.

## II. Observations and recommendations

12. The Workshop considered the importance of opportunities associated with the International Heliophysical Year 2007 for the participation of developing countries in activities recommended by the Committee on the Peaceful Uses of Outer Space and stressed the importance of early preparation for their possible participation in such activities.

13. The Workshop strongly recommended that the programme for the next United Nations/European Space Agency workshop on basic space science be organized for the benefit of developing countries and that it be structured so as to enable developing countries to participate actively in the associated scientific plans developed in the context of the International Heliophysical Year.

14. The Workshop noted with appreciation the invitations of the Russian Federation and the United Arab Emirates to host workshops on basic space science.

15. The Workshop expressed support for the ongoing COSPAR/IAU programme for education and training in basic space science at the professional level in developing countries. The continuation of the programme, which was a follow-up to past basic space science workshops, was supported. A better coordination of all programmes organized in different research areas by independent organizations could significantly enhance their impact.

16. The Workshop recommended examination of the feasibility of the creation of an independent source of funding, supported by interested parties, to facilitate the execution of global and regional basic space science studies, through small grants for active stimulation of multinational and cross-regional basic space science research initiatives.

17. The Workshop observed with satisfaction that cross-national and interregional initiatives were developing further, using basic space science facilities, established over the past decade. Formalization of networks and working groups with common goals with a view to further coordination of research work would be helpful for future acceleration of participation in such initiatives. Working groups could be especially beneficial to advance work in the following areas:

    (a) Selected variable star observations;

    (b) Study of comets, asteroids and near-Earth objects;

    (c) Application of non-extensive statistical mechanics to astronomical problems;

    (d) Sharing expertise with technical instrumentation;

    (e) Access to virtual observatories as developed in national environments.

18. The Workshop noted that the close collaboration between observatories in Indonesia, Malaysia and Paraguay would supply important continuous observing capability for the study of objects where long uninterrupted series of observations were essential for the understanding of objects such as delta scuti stars, Ap stars, dwarf novae and eclipsing binaries, which showed physical phenomena with periods of less than one day. Extending such collaborations to other observatories at different longitudes would make an important contribution to worldwide coverage of such phenomena.

19. The Workshop commended the NASA Astrophysical Data System (ADS) for its work and success in laying out and implementing road maps for the improved access for all scientists to the scientific literature, and expressed hope that the support for that work would be continued in the future. ADS was of prime importance for developing countries. The continued support of mirror site aspects of ADS was important and should be seriously considered in all countries where international boundaries crossing networks presented problems for the scientists.

20. The Workshop observed that the re-scanning of the historical literature by ADS was an important tool in equalizing access, in particular for scientists in developing countries, to knowledge in the basic space science accumulated over the past century.

21. The Workshop emphasized that various virtual observatory initiatives in a number of countries could contribute considerably to the accelerated development of basic space science in developing countries.

22. The Workshop emphasized that the ongoing interchange of standards and other necessary commonalities, one of the activities of the International Virtual Observatory Alliance (IVOA), would greatly enhance the value of individual virtual observatory initiatives.

23. The Workshop agreed that, even though important work had already been done for archival access to planetary data in a national context, further discussions between the planetary data archives community and the virtual observatory initiatives were recommended to ensure the incorporation of the large amounts of data currently being collected and foreseen for the future in a consistent way in the virtual observatory initiatives. This would be particularly important for data generated during future missions.

24. The Workshop encouraged close cooperation between ADS and virtual observatory initiatives that would open new perspectives for scientists from developing countries to participate at the forefront of new discoveries in basic space science.

25. The Workshop noted with satisfaction the continued installation of planetariums and telescopes in developing countries through the official development assistance programme of the Government of Japan, in particular its continuing support for Bolivia, Ethiopia and Pakistan.

26. The Workshop highlighted a number of hands-on web sites developed by major astronomical observatories and space agencies that were an important source of public information and educational material, as well as a source of motivation for public participation in basic space science, and were accessible to all countries. Professionals in basic space science should be aware that their participation in the facilitation of access to and public awareness of such resources was essential for optimization of their impact.

27. The Workshop noted with interest the opportunities offered by concurrent design for early interaction in the mission design for international space missions. This should be brought to the attention of decision and policy makers. A joint demonstration of the two currently most mature facilities, in operation at the Jet Propulsion Laboratory and ESA, to the Committee on the Peaceful Uses of Outer Space might be an important aid to future participation by developing countries in space projects.

28. The Workshop recommended plans for an outreach effort on the history of astronomy at the University of Sonora, Mexico. In that regard, information related to the cultural and human aspects of the history of astronomy should be forwarded to the University of Sonora (jsaucedo@cosmos.astro.uson.mx). The display of such information on a public access web site would highlight the cultural aspects of basic space science over the history of humankind as a whole.

## III. Summary of presentations

### A. Developing basic space science worldwide: a decade of United Nations/European Space Agency workshops

29. The first Workshop on Basic Space Science was held in 1991 and created a unique forum for scientific dialogue among scientists from

developing and industrialized nations. After more than a decade of annual workshops, the deliberations of the Twelfth Workshop brought together information on historical activities, the plans that had been developed over the past decade in different nations and the results that had materialized during that time in different developing and industrialized nations. Results addressed in the Workshop were the achievements of a truly international nature of all those involved in the previous workshops. Over time, mutual support from workshop participants has helped significantly to implement recommendations that emanated from the workshops. Workshop participants represented all economic regions of the world, specifically Africa, Asia and the Pacific, Europe, Latin America and the Caribbean, and Western Asia, which allowed recognition of the importance of a regional and at times a global approach to basic space science for developing and industrialized nations worldwide. The Workshop programme highlighted six specific scientific activities and investigations that had been carried out in various countries. The selection of the topics for the sessions of the Workshop was based on an assessment of proceedings and reports of past workshops containing scientific information presented in the workshops in the period from 1991 to 2002 and as contained in the decadal report on the workshops entitled "Developing basic space science worldwide: a decade of UN/ESA workshops".

## B. "Tripod" concept for accelerating the development of basic space science in developing countries

30. At the first Workshop, a concept was developed to promote basic space science in developing nations. That concept, which has come to be known as the "Tripod", comprises three elements. The first element is the provision of basic research tools at a level appropriate for a developing country, such as an astronomical telescope facility. Consequently, astronomical telescope facilities were established in Chile, Colombia, Egypt, Honduras, Jordan, Morocco, Paraguay, Peru, the Philippines, Sri Lanka and Uruguay. The second element was the development and provision of teaching material to allow the introduction of basic space science into established physics and mathematics curricula of universities in countries that implemented the Tripod concept. The third was the implementation of original research programmes in basic space science, at an appropriate level for the existing facilities and state of scientific development, such as variable star observing programmes supplemented by computer science, mathematics, physics and astronomy. Access to scientific literature and databases formed essential supplementary components to Tripod. The Workshop reviewed the progress in the implementation of Tripod and made appropriate recommendations for its future implementation in Bolivia, Ethiopia, Pakistan, the Syrian Arab Republic.

## C. Virtual observatories

31. State-of-the-art observing facilities on the ground and in space are producing large quantities of high quality data. These data are being captured in science archives with the goal of exploiting them in an optimum manner. The next logical step is to interconnect these archives in order to allow the users to retrieve the data in a simple and uniform way and to maximize the scientific use of these expensive resources. At the same time, it is useful to supply a suite of science visualization and analysis tools in order to facilitate the handling of the data. Funded by the European Commission and by the United States National Science Foundation, with contributions from major organizations such as ESA, NASA and the

European Southern Observatory, virtual observatory concepts are being developed in the United States and in Europe. On a smaller scale, virtual observatories are also being developed in other countries, such as China, India and the Russian Federation. To avoid redundancy, care is being taken to coordinate the efforts. This is being done through the international virtual observatory alliance, which also provides coordination with other virtual observatory activities worldwide. The Workshop reviewed ways and means of enabling developing countries to contribute to and benefit from virtual observatory activities.

### D. Astrophysical data system

32. ADS abstract service is a NASA-funded project that provides free Internet abstract search services. Currently, ADS has over 3.6 million references in the following four databases: (a) astronomy and planetary sciences; (b) physics and geophysics; (c) space instrumentation; and (d) astronomy pre-prints. Each database contains abstracts from hundreds of journals, publications, colloquiums, symposiums, proceedings, doctoral theses and NASA reports. ADS has 11 mirror sites in Argentina, Brazil, Chile, China, France, Germany, India, Japan, the Republic of Korea, the Russian Federation and the United Kingdom, which help to provide better global access to ADS. The workshops facilitated the use of ADS and its mirror sites in developing countries. The ADS article service provides free access to the full text of over 340,000 scientific papers published in astronomical journals, conference proceedings, newsletters, bulletins and books, amounting to 2.5 million scanned pages. More than 6.5 million links in the ADS system allow the user easy access to online data and other information related to the articles in ADS.

### E. Non-extensive statistical mechanics and astrophysics

33. A great variety of complex natural phenomena in many scientific fields exhibit power-law behaviour, reflecting a hierarchical or multifractal structure. Many of these phenomena seem to be susceptible to description and understanding using approaches drawn from thermodynamics or statistical mechanics, in particular approaches involving the maximization of entropy. In recent years, a large number of studies in many countries, including developing countries, has been devoted to a non-extensive generalization of entropy and of Boltzmann-Gibbs statistical mechanics and standard thermodynamics. This generalization has intrinsically non-linear features and yields power laws in a natural way. The Workshop addressed interdisciplinary applications of these ideas, in particular in the field of basic space science, as well as various phenomena that could possibly be quantitatively described in terms of these ideas.

### F. Concurrent design capability for the development of international space missions

34. During the Workshop, concurrent design capability, as available at the Jet Propulsion Laboratory and ESA, were employed for an interactive demonstration of the early design stages of international planetary missions. The demonstration established a remote videoconferencing link and a data link between Team X, located at the Jet Propulsion Laboratory, and the Workshop participants. The purpose of the demonstration was to establish a proof of the concept of concurrent and interactive mission design across international boundaries. This concept is an important step in establishing a capability for common mission design among NASA, ESA and the space agencies of other countries. The demonstration showed an international body of scientists, many from developing countries, attending the Workshop how a space mission concept was developed. Similar demonstrations had been made at the workshops in France (2000) and Argentina (2002).

### G. Lunar exploration

35. The Moon is currently the focus of international programmes of scientific investigation. Moon space missions currently operating or under development will lead to the future use of the Moon for science and technology development. The next step in human exploration beyond low-Earth orbit is to the Moon, the celestial neighbour closest to the Earth in the solar system. Many countries are developing lunar missions (China is developing the Chang'e mission, India the Chandrayan-1, Japan the Lunar-A and Selene missions and ESA the Small Missions for Advanced Research in Technology satellite 1 (SMART-1)) that offer opportunities for international cooperation. A number of justifications for an expanded lunar programme have been identified: in particular the assessment and use of potential ice and water resources at the lunar poles for human use and the development of energy resources for both the Moon and the Earth, as well as the establishment of lunar astrophysical observatories. For the future development of the Moon, the deposits of hydrogen indicated by the United States Clementine and Lunar Prospector missions must be understood to confirm their nature and importance for future planetary exploration, development, and human settlement. To encourage and stimulate the peaceful and progressive development of the Moon, national space agencies may operate and maintain an exploratory mission at a pole of the Moon to serve as a catalyst for future human missions within a decade.

## IV. International Heliophysical Year 2007

36. In 1957, a programme of international research, inspired by the International Polar Years of 1882-1883 and 1932-1933, was organized as the International Geophysical Year to study global phenomena of the Earth and geospace. The International Geophysical Year involved about 66,000 scientists from 60 countries, working from pole to pole at thousands of stations to obtain simultaneous, global observations on Earth and in space. The fiftieth anniversary of the International Geophysical Year will occur in 2007. It was proposed to organize in 2007 an international programme of scientific collaboration to commemorate the anniversary, to be called the International Heliophysical Year. Like the International Geophysical Year and the two previous International Polar Years, the scientific objective of the International Heliophysical Yearis to study phenomena on the largest possible scale, with simultaneous observations from a broad array of instruments. Today, unlike in previous international years, observations are routinely received from a vast array of sophisticated instruments in space that continuously monitor solar activity, the interplanetary medium and the Earth. These spacecraft, together with ground-level observations and atmospheric probes, provide an extraordinary view of the Sun and the heliosphere and the influence of both on the near-Earth environment. The International Heliophysical Year is a unique opportunity to study the coupled Sun-Earth system. Future basic space science workshops will focus on the preparations for the International Heliophysical Year worldwide, in particular taking into account the interests of and contributions from developing nations.

[i] See *Report of the Third United Nations Conference on the Exploration and Peaceful Uses of Outer Space, Vienna, 19-30 July 1999* (United Nations publication, Sales No. E.00.I.3), chap. I, resolution 1, para. 1 (e) (ii), and chap. II, para. 409 (d) (i).

[ii] *Official Records of the General Assembly, Fifty-sixth Session, Supplement No. 20* and corrigendum (A/56/20 and Corr.1), para. 74.